# A model of orbital angular momentum Li-Fi

Yuanying Zhang, Jikang Wang, Wuhong Zhang, Shuting Chen, and Lixiang Chen*

*Department of Physics, Jiujiang Research Institute and Collaborative Innovation Center for Optoelectronic Semiconductors and Efficient Devices, Xiamen University, Xiamen 361005, China*

Corresponding authors: *chenlx@xmu.edu.cn

Twisted light has recently gained enormous interest in communication systems. Thus far, twisted light has not yet been utilized for visible light communication to transmit data. Here, by exploiting the color and orbital angular momentum (OAM) degrees of freedom simultaneously, we construct a much higher-dimensional space spanned by their hybrid mode basis, which further increases the information capacity of twisted light. We build a new visible light communication system using a white light emitting diode, with red, green and blue (RGB) colors serving as independent channels and with OAM superposition states encoding the information. We connect our conceptually new RGB-OAM hybrid coding with the specially designed two-dimensional holographic gratings based on theta-modulation. After indoor free-space transmission, we decode the color information with an Xcube prism and subsequently decode the OAM superposition states with a pattern recognition method based on supervised machine learning. We succeed in demonstrating the transmission of color images and a piece of audio with the fidelity over 96%. Our point-to-point scheme with hybrid RGB-OAM encoding, not only increases significantly the information capacity of twisted light, but also offers additional security that supplements the traditional broadcasting visible light communications, e.g., Li-Fi.

# I. INTRODUCTION

It was Allen and coworkers who first recognized that a light beam with a helical phase front of $\exp(i\ell\varphi)$ carries an orbital angular momentum (OAM) of $\ell\hbar$ per photon, where $\ell$ is an integer, $\varphi$ is the azimuthal angle, and $\hbar$ is the Planck constant [1]. Recent years have witnessed a growing interest in its applications ranging from optical micromanipulation, optical imaging, to quantum information science [2-6]. Another particularly promising applications is the optical communication that employs OAM as the information carrier [7, 8]. This technique was proposed and developed under the circumstances that the bandwidth of laser communication proves to be insufficient to satisfy the exponentially growing demands in the foreseeable future. Light possesses several different degrees of freedom, e.g., the time, wavelength, and polarization state, all of which can be exploited to increase the channel capacity. Another potential solution to eventually cope with bandwidth issues is the space division multiplexing (SDM) [9], and in particular, the mode division multiplexing (MDM) [10]. In 2004, Gibson *et al.* first demonstrated the transfer of data information encoded by OAM modes to enhance the security of data transmission which was resistant to eavesdropping [11]. In the radio frequency (RF) regime, the generation and detection of RF OAM beam were proposed which could be chosen as data carriers in communication [12, 13], as the longer wavelength of a radio carrier wave is less sensitivity to various channel conditions and more divergence compared with an optical beam [14]. Also, Yan *et al.* demonstrated a 2.5m free-space link with a total capacity of 32 Gbit/s at a carrier frequency of 28GHz with eight coaxially propagating RF OAM beams [15]. The practical implementation of OAM multiplexing systems request new mode multiplexing and demultiplexing technologies in order to achieve high performance. In 2014, Wang *et al.* demonstrate the multiplexing and demultiplexing of information-carrying OAM beams for free-space data transmission with an achieved terabit data rate [16]. Also, a number of implementations of OAM-multiplexed transmission over relevant distances were demonstrated both optical fibers [17, 18]. Particularly important progress was made in 2014 by Krenn and coworkers who successfully implemented a 3km free-space communication with spatially modulated light through turbulence [19]. They further extended their scheme to work over a 143 km free-space established between two islands [20]. Very recently, Ren *et al.* also performed a high-speed classical communication with 400Gbit/s data rate by multiplexing OAM modes over 120 meters of free space in Los Angeles [21]. Here we add the color degree of freedom, e.g., the primary red, green and blue (RGB) colors, into the OAM eigenstates, and encode the information with such hybrid RGB-OAM mode base, and therefore, further increasing the information capacity of twisted light. In contrast with previous demonstration with commercial lasers, here we connect our hybrid RGB-OAM coding to the visible light communications with the use of a white light emitting diode (LED).

The visible light wireless communication can be traced back to the work in 1880 by Bell who transmitted voice data over 200m using sunlight [22]. Besides, several other demonstrations featuring fluorescent lights for communication at low data rates were investigated [23]. Nowadays, as an emerging technology for the optical wireless link, Light Fidelity (Li-Fi) represents a new paradigm for visible light communications [24]. In a Li-Fi system, the logical state of 1 and 0 is encoded by swiftly controlling the on and off state of LED, respectively, at a rate much faster than the response time of the human eye [25, 26]. This novel technology is supported by the ubiquitous and inexpensive LEDs. As mentioned before, OAM has been extensively employed in the laser communication systems, however, thus far no experimental demonstration that exploits OAM in the visible light communication has been conducted to transmit data with the white LED source. Here, we further increase the data capacity of twisted light by adding the color degree of freedom to the OAM eigenmodes such that we develop a new concept of hybrid encoding to further expand the OAM space. Specifically, by multiplexing and demultiplexing Red-Green-Blue (RGB) twisted light beams derived from LED, we demonstrate an effective visible light communication scheme with RGB colors serving as independent data channels while with OAM superpositions encoding the information. We accomplish the transmission of a color image of Albert Einstein, with the RGB channels naturally transmitting the RGB components of the image, respectively. We also transmit a piece of 8-bit Pachelbel's Canon in D based on Hexadecimal number sequence. After a 6-meter indoor free-space link, the color image and audio are recognized and decoded with a fairly good fidelity based on a supervised machine learning. The security of our white light communication can be additionally ensured by the inherent uncertainty of OAM states when possible eavesdropping occurs, which can serve as a point-to-point secure communication that supplements the traditional broadcasting Li-Fi.

## II. FRAMEWORK

### A. Hybrid encoding based on theta-modulation

Instead of a laser source, we use an inexpensive white LED as the source to produce the RGB twisted light beams based on theta-modulation [27, 28]. In case of mode division multiplexing (MDM), the orthogonal OAM eigenmodes have been directly employed in optical communications. But it seems that highly symmetric petal patterns of two opposite OAM superposition modes could be more advantageous, as they could be more efficiently detected and recognized in free space. We transmitted color image and audio by encoding 16 different OAM superposition modes, e.g., $\ell = 0, \pm 1, \pm 2, \ldots \pm 15$, where we choose the Laguerre–Gaussian (LG) modes with zero radial index $p = 0$ to represent these OAM states. The superposition of $\pm \ell$ OAM just gives rise to a $2\ell$-petal flower pattern of rotational symmetry, and the radius of the intensity patterns scale with $\ell$ [29]. By adding both the intensity and phase distribution of $\pm \ell$ OAM superposition modes to a linear grating, we obtain the 1st-order diffraction carrying the desired petal-like patterns. We aim to exploit the RGB model as three independent channels, so we are allowed to use the hybrid kets of $|\sigma, \pm \ell\rangle$ ($\sigma = R, G, B$) to label the OAM superposition states in the red, green and blue channels, respectively. Thus with such hybrid RGB-OAM modes we are able to construct a much higher-dimensional space beyond the OAM consideration only. In our demonstration below, for color image transmission, we use the color degree of freedom to encode the red, green and blue components of the image and OAM superpositions to encode the grayscale of each pixel, while for audio transmission, we use color to encode the sub-streams of audio data and OAM superpositions to encode the audio amplitude values. Our hybrid encoding strategy can be realized with a single spatial light modulator (SLM) based on the technique of theta-modulation. The specially designed hologram is generated by integrating three one-dimensional gratings with 0°, 60° and 120° orientations, with each one-dimensional grating being programmed to generate the desired OAM superposition modes individually. By compositing three one-dimensional grating of 0°, 60° and 120° orientations masks, we can obtain a two-dimensional holograms serving as the multiplexing phase mask. According to theta-modulation, we know that with a broadband LED illumination, three diffraction patterns will disperse spectrally and separate angularly in the Fourier plane, situating on the direction at 0°, 60° and 120°, respectively. A tailed-made opaque aperture is inserted at the Fourier plane, where three punctured pinholes are carefully adjusted at each diffraction to filter out the primary RGB components, with one pinhole in each 1st-order diffraction. After a telescope, the RGB light beams encoded with various OAM superposition modes can be produced and transmitted coaxially through free space towards the receiver.

For color images, we decompose the image based on the RGB color model. And naturally, the primary red, green and blue components are transmitted through the red, green and blue channels offered by the white LED, respectively. In each channel, the monochromatic image is sent pixel by pixel, with the gray scale of each pixel being encoded by the 16-level OAM superposition modes, specifically, $\ell = 0, \pm1, \pm2, ... \pm15$. For audio, its waveform describes a depiction of the pattern of sound pressure variation or amplitude in the time domain. We extract the discrete amplitude values between 0 and 255, and convert each value to two hexadecimal numbers, e.g., 0, 1, 2, 3…, e, f. In a similar way, we encode these 16 different hexadecimal numbers with OAM superposition modes of $\ell = 0, \pm1, \pm2, ... \pm15$, respectively. In order to increase the data rate, we divide the stream of hexadecimal numbers into three sub-streams, which are transmitted through the RGB channels, respectively.

### B. Decoding based on machine learning

At the receiver, two lenses of different focused lengths ($f_3$=150*mm* and $f_4$=50*mm*), forming a demagnification 4f system, collects the incoming RGB twisted light beams. As the information is RGB-OAM hybrid encoded, we first direct the incoming light to a Cross Dichroic X-cube Prism. This X-cube prism is a RGB combiner /splitter, which serves as a demultiplexer for the RGB channel signals. After coming out from the three ports of the X-cube prism, the red, green and blue signals are separated and then recorded by a color CCD camera (Thorlabs, DCU224C), respectively.

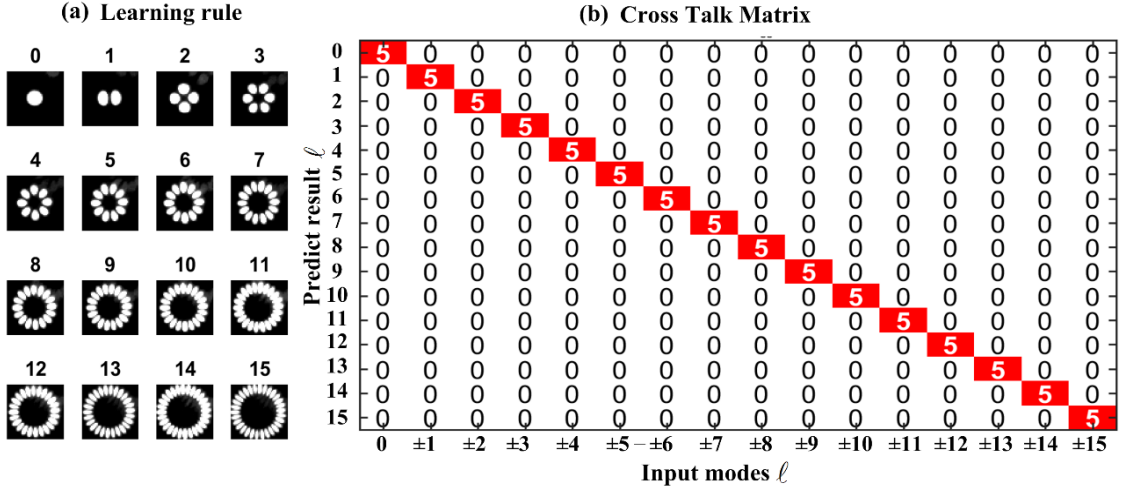

FIG. 1. Pattern recognition with unsupervised machine learning. (a) Learning rules: the OAM superposition modes, $\ell=0,\pm1,\pm2,...\pm15$, are assigned by the numbers, 0, 1, 2, ... , 15, respectively. (b) The validation crosstalk matrix of the predictor.

The next key step is to distinguish the sequential OAM superposition modes encoded in each color channel. We accomplish this by using a pattern classification recognition based on bagged classification trees with the bootstrap aggregating (acronym as bagging) method [30, 31], which is a kind of supervised machine learning. It accomplishes a task to infer a predictor from labeled training data, and the predictor can be used for mapping new examples. In our work, we use the successive OAM superposition modes to serve as original data. The input patterns of the OAM superposition modes from $\ell=0$ to $\ell=\pm15$ are designated as the numbers from 0 to 15 accordingly, as illustrated in Fig. 1(a). Firstly, we randomly partition the data into two parts, 90% of the image data as a training set for generating predictor, and the rest of the data as the validation set for evaluating the quality of the predictor. We category the 16 OAM superposition modes with the bagging classification method, which generates multiple versions of a predictor and uses them to get an aggregated predictor. Therefore, based on the bagging predictor, an aggregated predictor can be generated to predict a numerical outcome and does a plurality vote when predicting a class. Secondly, for testing the quality of the generated predictor, we use the predictor to recognize the mode in the validation set. Fig. 1(b) is the evaluation outcome, we notice that the predictor model can be generated and autonomously categorizes the 16 kinds of input mode structures, outputting a series of mode numbers. The crosstalk matrix represents the distinguishability of the input and measured OAM superposition which is dominantly diagonal. The results show that the predictor has good performance and can well distinguish 16 different OAM superposition modes. After this initialization, the predictor model can be applied to analyze the real image data.

**C. The bagged classification trees in machine learning**

The core of the pattern classification recognition is the use of bagged classification trees to get a predictor which can recognize different OAM superposition modes. In order to obtain the predictor, we input a set of learning materials for training. A learning set of $\{L\}$ consists of data $\{(y_n, x_n), n=1....,N\}$, where $y's$ are either class labels or a numerical response. Assuming a predictor can be defined by $\varphi(x,L)$, if the input is $x$ we can predict $y$ by $\varphi(x,L)$. In the learning process, the multiple samples are formed by making bootstrap [32] replicate of learning set and using these as new learning sets $\{L_K\}$, each consisting of $N$ independent observations. In our work, the training process starts from getting 50 sets of 16 OAM modes superposition. Here the OAM superposition patterns and the $\ell$-values are corresponding to the input $x$ and the numerical response $y$, respectively. Specifically, the input OAM superposition modes from $\ell = 0$ to $\ell = \pm 15$ are designated as the numbers from 0 to 15, respectively, as shown in Fig. 1(a).

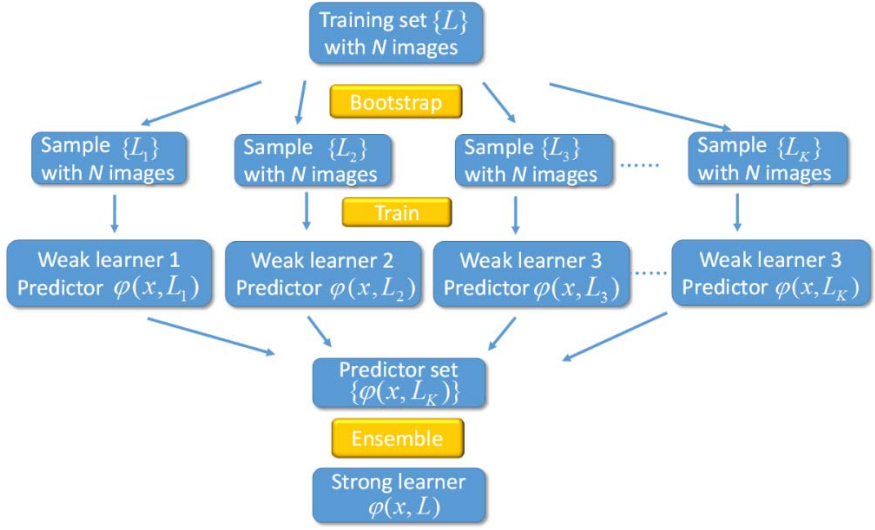

FIG. 2. Strategy of pattern recognition using bagged classification trees.

The predictor to identify the OAM superposition pattern is derived by utilizing an ensemble of bagged classification trees, and the strategy is schematically illustrated in Fig. 2. We randomly partition the data into two parts, 90% of the image data serve as the training set for generating predictor, and the rest serve as the validation set for evaluating the performance of the model. In order to get a good predictor, the machine learning process carries out three main steps [33-35]: bootstrap sampling, train, and ensemble, see Fig. 2. Firstly, we input the training set $\{L\}$ which consists of $N$ images, here $N = 720$. Secondly, by the method of bootstrap sampling, it randomly samples with replacement and generates a series of sample $\{L_1\},\{L_2\}...\{L_K\}$ from the training set. Here we aim to use the sample set to get a better predictor than a single learning set predictor. For every sample we use a weak learner, i.e., the decision tree to classify independently in every region, generating a predictor set $\{\varphi(x, L_K)\}$ which consist of $K$ predictors. Finally, according to the ensemble learning strategy, a strong learner $\varphi(x,L)$ can be obtained by averaging and voting all the $K$ predictors $\{\varphi(x, L_K)\}$. Thus, with the predictor $\varphi(x,L)$, if an arbitrary OAM mode-intensity image is input, we can obtain a predicted result of $\ell$-value accordingly. We also apply the predictor model to classify the test data, and we show in Fig. 1(b) the testing result of validation set, from which the good performance can be seen clearly.

## III. RESULTS

Our optical system of OAM-based visible light communication is shown in Fig. 3. The white light derived from a 125 mW LED (Daheng, GCI-060411) with a bandwidth ranging from 440nm to 670nm. A telescope is used to collimate the white-light beam, which then illuminates the SLM uniformly. It is noted that the usage of telescoping and beam expansion will alleviate the error in waist sizes and strong curvature [36]. The sender is basically a 4*f* system, consisting of two lenses with focal length $f_1$=300*mm* and $f_2$=1000*mm*, that realizes the theta-modulation to generate the colorful OAM superposition states. The SLM (Hamamatsu, X13138-01) is a phase-only modulator, which is capable of operating in a wide range of the spectrum as well as broad band source, allowing us to generate customized digital hologram to encode the information [37, 38]. After transmitting in free space, the red, green and blue channels will be demultiplexed by an X-cube prism at the receiver, and then recorded by a color CCD camera, respectively, see Fig. 3(b). The 16 different OAM superposition modes are identified with pattern recognition, based on the aforementioned method of machine learning, thereby decoding the information.

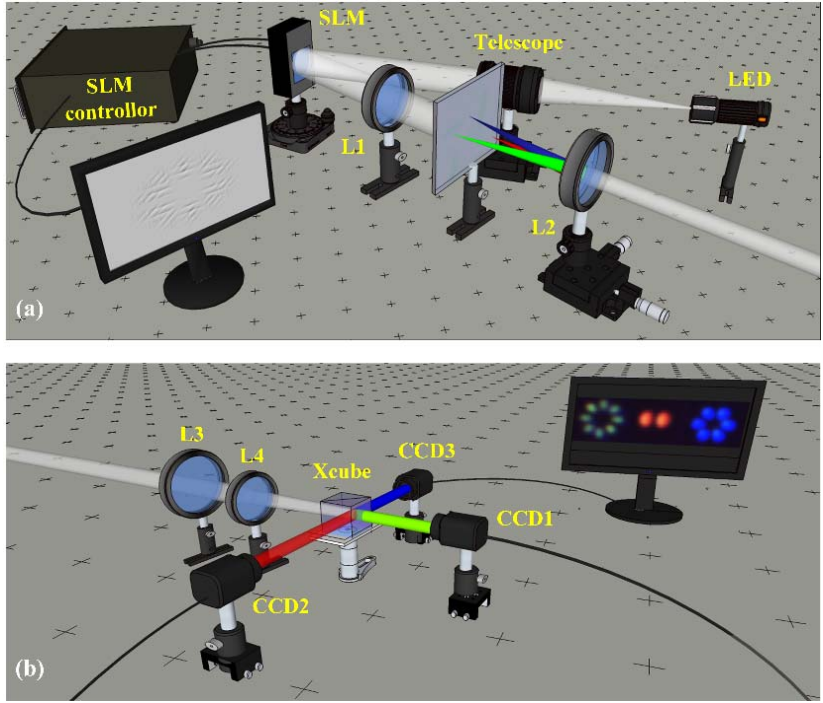

FIG. 3. Sketch of the experimental setup for visible light communication link based on the RBG twisted light encoding/decoding. (a) The sender, (b) The receiver, see the text for details.

### A. Transmission of the primary RGB colors

To verify the validity of our communication system, we first demonstrate the transmission of a simple tri-circle image of three RGB primary colors. The RGB color model is an additive color model in which red, green and blue light are added together in various ways to reproduce a broad array of colors. As shown in Fig. 4(a), a secondary color is formed by the sum of two primary colors of equal intensity, e.g., green and blue make cyan, red and blue make magenta, while red and green make yellow. By adding all three primary colors together, it yields white. Generally, a color is expressed as an RGB triplet $(r,g,b)$, which indicates how much of each of the red, green and blue is included. For the tri-circle image of Fig. 4(a), it has a resolution of 200×184=36800 pixels and a 4-bit grayscale value of the RGB triplet $(r, g, b)$. To transmit the full color image, a natural and effective way is to use the red, green and blue colors of white LED as three independent channels to send the red, green and blue components of the image, respectively. And for each channel, we further use the 16 different OAM superposition modes to encode the 4-bit pixel grayscale values, thereby creating a basis set of 48 modes encoded in both color and OAM degrees of freedom. This is the key for our encoding technique and for achieving a higher channel capacity in the implementation of the optical link. Hence, the 200×184=36800 tri-circle image can be mathematically converted to three data streams, each with 36800 sequence numbers.

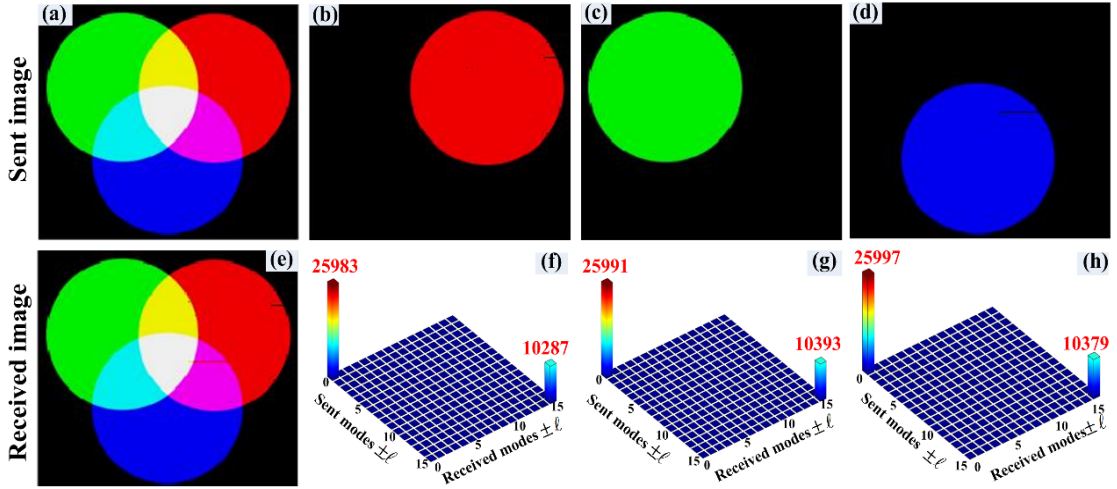

FIG. 4. Experimental results for transmission of a tri-circle image of RGB primary colors. (a) The 4-bit original image to be sent. (b-d) The received and reconstructed red, green and blue components of the image, respectively. (e) The full-color image reconstructed by adding three primary components (b-d) together. (f-h) The corresponding crosstalk matrices of OAM superposition modes for red, green and blue channels, respectively, where the numbers denote the events and zero OAM modes are used to transmit the dark background trivially.

Based on our above encoding strategy, we can rapidly load the multiplexing masks on SLM and the grayscale values of red, green and blue components are sent pixel by pixel. At the receiver, the red, green and blue colors are directly separated and then demultiplexed by the X-cube prism. Three color CCD cameras placed at the three output ports of the X-cube prism record the successive patterns in the video form simultaneously. Based on the decoding strategy, we extract each frame of the videos and decode the OAM superposition modes by means of the aforementioned pattern recognition with machine learning. Thus we can reconstruct the red, green and blue components of the received image, as shown in Fig. 4(b), 4(c) and 4(d), respectively. As the image is composed of three circles, each with a primary color, the received image after reconstruction is merely a circle of the corresponding primary color. By adding these red, green and blue components together, we finally reconstruct the received image of Fig. 4(e). For a quantitative estimation, we also plot the crosstalk matrices of OAM superposition modes for the RGB channels in Fig. 4(f), 4(g) and 4(g), respectively. By defining the error rate as the ratio between the wrong bits and all the bits, we estimate that the pattern recognition predictor distinguishes the mode patterns with a relatively low error rate, e.g., 0.3% for red, 0.11% for green, and 0.08% for blue channels, respectively. While for the whole image of Fig. 4(e), the average error rate is calculated to be around 0.17%, therefore, showing the favorable performance of our visible light communication system.

### B. Transmission of a color Albert Einstein image

More generally, we performed another experiment to transmit a color Albert Einstein image of Fig. 5(a), which has a 200×164=32800 pixels and exhibits a more complicated color mixture. Again, the RGB components of the Albert Einstein image are sent, naturally, through the RGB channels of the LED light, respectively. At the sender, we encode and transmit each RGB pixel according to the same strategy as above. At the receiver, we record and analyze the videos recorded in each RGB channel, and decode the grayscale information pixel by pixel. The recovered RGB images were shown in Fig. 5(b), 5(c) and 5(d), respectively. And based on the measured crosstalk matrices in Fig. 5(f), 5(g) and 5(h), we calculate the error rates to be 0.14%, 4.2%, and 3.51%, respectively. Besides, we can see that the most frequently encoded OAM superposition modes are $\ell=\pm 14$, $\pm 6$ and $\pm 3$, which are used 3271, 3211 and 5181 times in the red, green and blue channels, respectively. Finally, we recover the original image successfully with a high fidelity, see Fig. 5(e), whose average error is only around 2.62%. These errors occur mainly due to the misrecognition of high-order mode patterns, as they possess a complicated structure of dense petals.

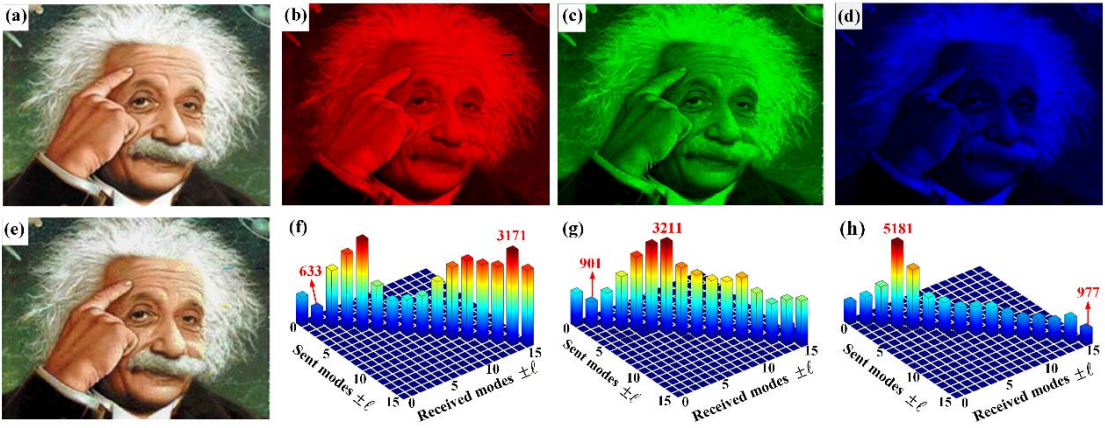

FIG. 5. Experimental results for transmission of the color Albert Einstein image. (a) The 4-bit original image to be sent. (b-d) The received and reconstructed red, green and blue components of the image, respectively. (e) The full-color image reconstructed by adding three primary components (b-d) together. (f-h) The corresponding crosstalk matrices for red, green and blue channels, respectively.

## C. Transmission of a piece of Pachelbel's Canon in D

To demonstrate the versatility of our scheme, we further transmit the Canon in D composed by Johann Pachelbel with the sampling frequency of 8000Hz and 8-bit depth. For audio, its waveform describes a depiction of the pattern of sound pressure variation or amplitude in the time domain, see Fig. 6(a). As computers don't store sound, instead, they store math, we need the conversion of a sound wave (a continuous signal) to a sequence of samples (a discrete-time signal), e.g., a series of 1s and 0s [39]. The sample rate of 8000Hz describes how fast the computer is taking those "snapshots" of sound, namely, 8000 samples are obtained per second on average. Each sample of an audio signal must be ascribed a numerical value to be stored and processed by the computer. For example, the 8-bit depth describes the number of bits of information in each sample, namely, the audio amplitude values range from 0 to 255. Here we aim to transmit a small piece of 20.862s audio through our OAM visible light communication system. For this, we extract the 8-bit amplitude values and convert each value to two hexadecimal numbers, (0, 1, 2, 3…, e, f). Hence, the audio can be mathematically represented by a sequence of 166,894 hexadecimal numbers. In our scheme, we just use the OAM superposition modes to encode the hexadecimal number. Thus the audio information can be transmitted through the free-space link after the spatial mode encoding with SLM. To further increase the transmission rate by taking full advantage of the three RGB independent channels, we divide the whole stream of the hexadecimal numbers into three sub-streams and send them through the red, green and blue channels, respectively, each with 55631 hexadecimal numbers. At the receiver, in a similar way, we first separate the RGB channels by using the X-cube prism, and decode the OAM superposition modes with the same method of pattern recognition. Then the data sequences in the RGB channels are all recovered, thus we reconstruct the piece of the music finally, as was shown in Fig. 6(b). If we define the fidelity as the ratio between the number of the correct amplitude values in the reconstructed audio and the total number of amplitude values in the original one, we can reach a relatively high fidelity over 96%.

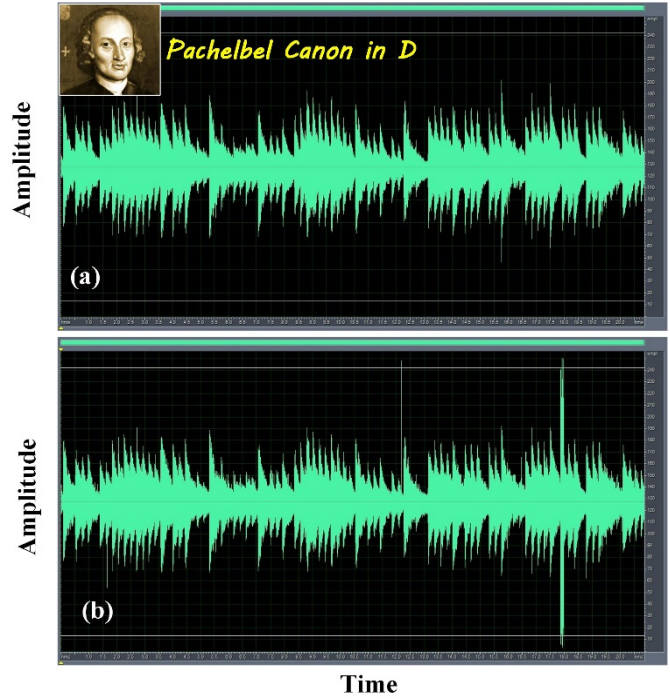

FIG. 6. The waveform graphs of 20.862s Canon in D composed by Johann Pachelbel. (a) The original audio waveform. (b) The received and recovered waveform by our OAM-based visible light communication system, see [40] for the Audio S1 for (a) and Audio S2 for (b).

## IV. CONCLUSION AND OUTLOOKS

In the above proof-of-principle experimental demonstration, we have shown the potential of twisted light in visible light communication. We are now facing an imminent shortage of radio frequency (RF) spectrum, i.e., the spectral efficiency (the number of bits successfully transmitted per Hertz bandwidth) of wireless networks has become saturated. Li-Fi, the high-speed communication and networking variant of visible light communication, aims to unlock a vast amount of unused electromagnetic spectrum in the visible light region [41]. Although, here, only the red, green and blue components of the white LED are utilized, we anticipate that our scheme could become a viable technique soon to help mitigate RF communications spectrum bottlenecks, with the visible spectrum being fully explored.

The data transmission rate of our present system is mainly restricted by the response time of SLM we use in the optical system, which has a limited frame rate of 60 Hz only. Further technology for speeding up the modulation of light's spatial structure will improve the transmission rate dramatically. For instance, high-performance digital mirror device (DMD) possesses the ability to rapidly switch different modes at a frame rate as high as 20 kHz. Also, further improvements such as high-speed integrated OAM transmitters might lead to the use of OAM superposition modes as an effective and fast way to encode information [42]. Besides, we can see that the transmission capacity may be further increased if the radial degree of freedom in OAM beams [43, 44] as well as vector beam [45] are considered. With these techniques incorporated, our scheme will become more promising for future practical communications. On the other hand, security is another important factor for communication systems. Unlike the broadcasting style of a traditional Li-Fi, our transmission link is an indoor point-to-point model, therefore, offering an additional security as is ensured by OAM measurement. It was shown that the information encoded in the OAM basis is resistant to possible eavesdropping, as any attempt to sample the beam away from its axis will be subject to an angular restriction and a lateral offset, both of which result in inherent uncertainty in the measurement [11].

In summary, we have proposed a hybrid RGB-OAM encoding/decoding strategy in a high-dimensional space and demonstrated its application in visible light communication system to further increase the data capacity of twisted light. The sender uses the theta-modulation to realize the hybrid encoding and transmit the information through the red, green and blue free-space channels offered by a white LED. The receiver demultiplexes and decodes the red, green and blue OAM superposition modes based on a single Xcube prism and the supervised machine learning. We have succeeded in transmitting both the color images and a piece of audio with a relatively high accuracy over 96%. By comparing with the traditional OAM communication with a commercial laser, our scheme exploits the additional color degree of freedom that enables the information capacity of twisted light. Besides, our point-to-point scheme offers additional security that can supplements the traditional broadcasting Li-Fi. Finally, we would like to point out that our work has brought up many interesting questions, some of which we discuss here, and which need to be fully addressed before our ideas can be realized in a commercial communication system.

## ACKNOWLEDGMENTS

We would like to thank Mr. Xiaochuan Jiang for Mathlab codes. This work is supported by the National Natural Science Foundation of China (NSFC) (11104233, 11474238), the Fundamental Research Funds for the Central Universities (2011121043, 2012121015), the Fujian Province Funds for Distinguished Young Scientists, the Program for New Century Excellent Talents in Fujian Province University, and the program for New Century Excellent Talents in University of China (NCET-13-0495).